\documentclass[10pt,letterpaper,twocolumn]{article} 

\usepackage{ol2}
\usepackage[draft]{hyperref}
\usepackage{amsmath}

\begin{document}
\twocolumn[
\title{Passive optical switching of photon pairs using a spontaneous parametric fiber loop}
\author{Ningbo Zhao, Lei Yang, Xiaoying Li$^{*}$}
\address{College of Precision Instrument and
Opto-electronics Engineering, Tianjin University, Key Laboratory of Optoelectronics Information Technology, Ministry of Education, Tianjin, 300072,
P. R. China}


\begin{abstract}
We study a novel scheme named spontaneous parametric fiber loop (SPFL), configured by deliberately introducing dispersive elements into the nonlinear Sagnac loop, and show it can function as a passive switch of photon pairs. The two-photon state coming out of SPFL highly depends on the dispersion induced phase difference of photon pairs counter-propagating in the loop. By properly managing the dispersive elements, the signal and idler photons of a pair with a certain detuning and bandwidth can be directed to the desired spatial modes of SPFL. If the photon pairs are used to generate heralded single photons, the SPFL can be viewed as a switch of single photons. Moreover, our investigation about the dispersion based phase modulation is also beneficial for designing all fiber sources of entangled photon pairs.

\end{abstract}


] 

\noindent

There has been a lot of interest in optical switching schemes~\cite{El-Bawab}. For quantum communication exploiting the fiber network, all fiber switching devices actuated by a single photon are desirable~\cite{Elliott02,Hall11}. Current fiber-based quantum switches tend to control a weak beam with a strong one by using the Kerr-phase modulation~\cite{Hall11}. In this letter, we study a scheme named spontaneous parametric fiber loop (SPFL), configured by deliberately introducing dispersive elements into the nonlinear Sagnac loop, and show it can function as a switch of photon pairs by using dispersion based phase modulation. The two-photon state coming out of SPFL highly depends on the dispersion induced phase difference of photon pairs counter-propagating in the loop. By properly managing the dispersive elements, signal and idler photons of a pair with a certain detuning and bandwidth can be directed to the desired spatial modes of SPFL.

As shown in the square frame in Fig. 1(a), the SPFL consists of a piece of nonlinear fiber (NLF), two pieces of standard single mode fibers (SMFs), SMF1 and SMF2, a fiber polarization controller (FPC), and a 50/50 fiber coupler (FC). The NLF serves as the nonlinear medium of spontaneous four wave mixing (SFWM), in which two
pump photons at frequencies $\omega _{p1}$ and $\omega _{p2}$ scatter through the $\chi_{xxxx} ^{(3)}$
nonlinearity to to create
energy-time entangled
signal and idler photons at frequencies $\omega _s$ and $\omega _i$, respectively, such that $\omega _{p1}+\omega _{p2}=\omega _s+\omega
_i$. While SMF1 and SMF2 act as the dispersive elements of photon pairs.
Each linearly polarized pump field, $E_{pi}$ (i=1,2), injected into the SPFL is split into two pump waves traversing in a
counter-propagating manner by the 50/50 FC, and
the two pump fields in the clockwise (CW) and counter-clockwise (CCW) directions produce co-polarized photon pairs, $\left| \omega _s,\omega _i\right\rangle_c$ and $\left| \omega
_s,\omega _i\right\rangle_d$, respectively, where the footnotes $c$ and $d$ denote the propagation direction.
The SPFL is then preceded by a
circulator (Cir), which redirects the SPFL reflected photons to a separate spatial mode. For clarity, we label the two output
modes of SPFL as "$a$" and "$b$".

\begin{figure}[htb]
\includegraphics[width=7.5cm]{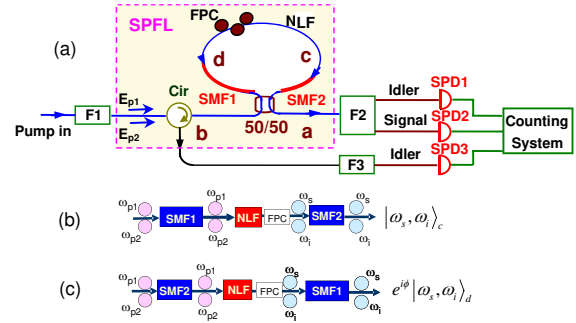}
\caption{(Color online) (a) Experimental setup. The configuration of SPFL is shown in the square frame. The generation and transmission of photon pairs in the CW and CCW directions of SPFL are illustrated in plots (b) and (c), respectively.}
\end{figure}

We note that the SPFL is a bi-directional amplifier, in the sense that the vacuum injected from either input port can be amplified. The sum of the phases of the signal and idler photons of a pair equals to that of the two pump photons, but the phase of individual signal or idler photons is random. Therefore, the intensities of individual signal and idler photons in mode $a$ and $b$ are the same. This is different from the parametric loop mirror reported in Ref. \cite{Mori95}, in which both the signal wave and idler wave originated from the amplification of injected signal via four wave mixing have well defined phases.

For each classical pump $E_{pi}$, the SPFL acts as a fiber loop reflector~\cite{Mortimore88}. When the pump power reflected back to mode "b", $I_{pi}$, takes the extreme values, the polarization modes of the counter-propagating pumps are automatically matched; otherwise, $I_{pi}$ itself only is not related to the mode matching~\cite{li09}. If the polarization of the pump fields in SPFL is carefully matched, we have $I_{pi}\propto\cos^2(\phi_{pi}/2)$, where the phase difference of the counter-propagating waves $\phi_{pi}$ ($i=1,2$) is originated from the different polarization transformation and can be adjusted by the FPC. Therefore, when the loop functions as an high reflector (HT) or high transmitter, we have $\phi_{pi}=0$ or $\phi_{pi}=\pi$.

To explain why SMF1 and SMF2 are referred to as the dispersive elements of photon pairs, let's analyze the generation and transmission of the photon pairs in SPFL. As illustrated in Fig. 1(b) and 1(c), when the mode is well matched (the same in polarization) by adjusting FPC, in the CW (CCW) direction, two pump photons with a phase delay introduced by SMF1 (SMF2) are scattered into a pair of signal and idler photons in NLF, and then the photon pairs experience phase delay in propagating through SMF2 (SMF1). Before coming out of SPFL, the counter-propagating photon pairs with a phase difference $\phi$ are then recombined at the 50/50 FC. Since the signal and idler photon pairs inherit the phase of the two pump photons, we have
$\phi=\phi_{p1}+\phi_{p2}+\phi_d$
with $\phi_d=(k_{p1}'+k_{p2}'-k_s'-k_i')L_2-(k_{p1}+k_{p2}-k_s-k_i)L_1$ denoting the dispersion induced phase difference, where $k_{pi}$ ($k_{pi}'$) (i=1,2), $k_{s}$ ($k_{s}'$), and $k_{i}$ ($k_{i}'$) are the wave-vectors of pump $E_{pi}$, signal and idler fields in SMF1 (SMF2), respectively, and $L_1$ ($L_2$) is the length of SMF1 (SMF2). Accordingly,
the output state of the SPFL is~\cite{li09}
\begin{eqnarray}
\left| \psi\right\rangle =\cos{\frac{\phi}2}\left| \psi_1\right\rangle +\sin{\frac{\phi}2} \left| \psi_2\right\rangle,
\label{state-1}
\end{eqnarray}
where $\left| \psi_1\right\rangle =(\left| \omega _s,\omega _i\right\rangle
_a+\left| \omega _s,\omega _i\right\rangle _b)/\sqrt{2}$,
$\left| \psi_2\right\rangle =(\left| \omega _s\right\rangle _a\left|
\omega _i\right\rangle _b+\left| \omega _i\right\rangle _a\left| \omega
_s\right\rangle _b)/\sqrt{2}$, and the footnote $a$ and $b$ denote the spatial modes of signal and idler photons~\cite{Chen07b,li09}. Particularly, we can obtain the state $|\psi\rangle=\left| \psi_1\right\rangle $
for $\phi =2n \pi$, and $|\psi\rangle=\left| \psi_2\right\rangle $
for $\phi =(2n+1)\pi$, where $n$ is an integer.
In the former case, both signal and idler photons of a pair come out in the same spatial mode, they are both simultaneously in either mode $a$ or mode $b$; in the latter case, the simultaneously created signal and idler photons have different spatial modes, i.e., if
one photon in mode $a$ is known to be in frequency $\omega _i$ then the other
one in mode $b$ is determined to have frequency $\omega _s$, or vice versa.


Since the procedure of matching polarization mode in SPFL is complicated for $\phi_{pi}\neq0$ or $\phi_{pi}\neq\pi$, we set the SPFL functions as an HR for pump to ensure the modes of photon pairs are automatically matched~\cite{Mortimore88,li09}. In this situation, we have $\phi_{p1}+\phi_{p2}=0$ and $\phi=\phi_d$. Therefore, Eq. [\ref{state-1}] can be written as
\begin{eqnarray}
\left| \psi\right\rangle =\cos{\frac{\phi_d}2}\left| \psi_1\right\rangle +\sin{\frac{\phi_d}2} \left| \psi_2\right\rangle.
\label{state}
\end{eqnarray}
For such a state, the single-count probability of the signal and idler photons in each spatial mode is constant, while the coincidence-count probability of signal and idler photons in same and different spatial modes is given by
\begin{eqnarray}
C_{si} \propto1\pm\cos \phi_d.
\label{coin}
\end{eqnarray}
Moreover, if the dispersion property of SMF1 and SMF2 is the same, we get
\begin{eqnarray}
\phi_d\approx-\Delta \Omega ^2 \beta_2(L_1-L_2),
\label{dpm}
\end{eqnarray}
where $\Delta \Omega$ is the detuning of photon pairs, and $\beta_2$ is the second-order dispersion coefficient of SMFs. For the non-degenerate photon pairs ($\omega_{s}\neq\omega_{i}$) produced by one  pulsed pump ($\omega_{p_1}=\omega_{p_2}=\omega_{p}$), we have $\Delta \Omega=\omega_s-\omega_p=\omega_p-\omega_i$; while for the degenerate photon pairs  ($\omega_{s}=\omega_{i}=\omega_{si}$) produce by two pulsed pumps ($\omega_{p1}\neq\omega_{p2}$), we arrive at $\Delta \Omega=\omega_{si}-\omega_{p1}=\omega_{p2}-\omega_{si}$.

Equations [\ref{state}]-[\ref{dpm}] show that the signal and idler photons of a pair can be directed to the required spatial modes of SPFL by properly managing the dispersion property and lengths of SMF1 and SMF2. Based on this feature, the SPFL can be viewed as a passive switch of photon pairs. Moreover, it is worth to point out that the switch only functions for the photon pairs with narrow bandwidth and with detuning satisfying a certain relation. For example, to successfully route the photon pairs to different spatial modes, the condition $\Delta \Omega =\sqrt{|(2n+1)\pi/(\beta_2(L_2-L_1)|}$ should be fulfilled.

Our experimental setup for testing the SPFL is shown in Fig. 1(a). The NLF is 300 m dispersion shifted fiber (DSF) with zero-dispersion wavelength of about $1547$\,nm, and the lengths of SMF1 and SMF2 (G.652) are about $3$ m and $1$ m, respectively. The linearly polarized pulsed pump is spectrally
carved out from a mode-locked femto-second fiber laser. Photons at the signal and idler
wavelengths from the fiber laser that leak through the spectral carving
optics are suppressed by passing the pump pulses
through a band-pass filter
F1 with FWHM of 0.9 nm. The pulse width and central wavelength of pump launched into the SPFL are about $4$\,ps and $1547.5$\,nm, respectively.  Under this condition, an efficient co-polarized SFWM with broad band phase matching is realized in DSF. In order to ensure the condition $\phi=\phi_d$ is fulfilled, we adjust the FPC to maximize the pump power reflected back to the port "b".

To reliably detect the photon pairs, an
isolation between the pump and signal/idler photons in excess of
100\,dB is required. In mode $a$, we achieve
this by first exploiting the mirror like property of the SPFL, which provides a pump rejection
greater than 30 dB, then sending the photons through a dual band filter
F$2$ with a pump rejection in excess of 75 dB~\cite{Fiorentino02}. The central wavelengths of F$2$ in both signal and idler band are tunable, and the FWHM of F$2$ in each band is 0.7 nm. In mode $b$, the idler or signal photons are selected by a tunable filter F3 with pump rejection and FWHM of $\sim$120 dB and 1.3 nm, respectively.
The idler and signal photons propagating through F2 are recorded by single photon detectors (SPD, id200), SPD1 and SPD2, respectively. The signal or idler photons passing through F3 are counted by SPD3.
The SPDs are operated in the gated Geiger mode. The $2.5$\,ns
gate pulses arrive at a rate of about 3.1 MHz, which is $1/8$ of the
repetition rate of the pump pulses, and the dead time of the gate
is set to be 10 $\mu$s.

\begin{figure}[htb]
\includegraphics[width=6.5cm]{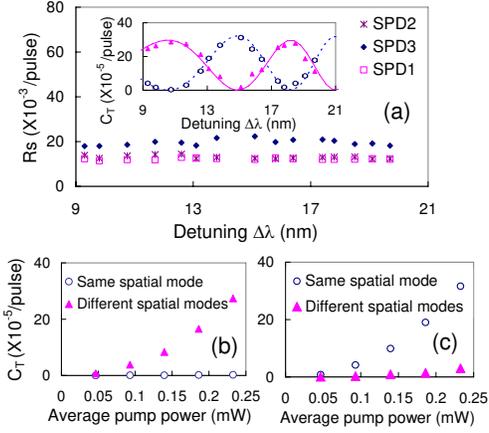}
\caption{(Color online) Experimental results. (a) Single count rate of each SPD, Rs, vs the detuning $\Delta \lambda$. The inset is the true coincidence $C_T$ vs $\Delta \lambda$. The hollow circles and triangles represent the data for photon pairs in the same and different spatial modes, respectively. The solid and dashed curves are the fits to Eq.(5) with $\alpha=0.0435$ ps$^2$ and with $\xi=29.5$ and $\xi=32.3$, respectively. (b) and (c) plot $C_T$ vs the average pump power for $\Delta \lambda =10.75$ and $\Delta \lambda =15.2$ nm, respectively.}
\end{figure}

We first test the dependence of the output state $\left| \psi\right\rangle$ upon the dispersion induced wavelength dependent phase modulation [Eq. (2)]. Using Eq. (4), and taking the broadband nature of pump, signal and idler fields into account, the true coincidence rate of signal and idler photon pairs in the same and different spatial modes is
\begin{eqnarray}
C_T = \xi[1\pm\cos (-4\pi^2 \alpha\Delta \lambda^2 c^2/(\lambda _{i0}\lambda _{p0})^2) ],
\end{eqnarray}
where $\alpha=\beta_2(L_2-L_1)$, $c$ is the speed of light in vacuum, $\Delta \lambda$ is the detuning in wavelength space, and the coefficient $\xi$ is determined by the detected counting rate of individual signal photons via SFWM, the total detection efficiencies in idler band, and the bandwidths of signal and idler photons~\cite{li10}.
In the experiment, the pump power is fixed at 0.23 mW, the central wavelength of F3 is in the idler band, and the detuning $\Delta \lambda = \lambda_{p0}-\lambda_{i0}$ is varied by changing the central wavelengths of F2 and F3, where $\lambda_{i0}$ is the central wavelength of idler photons. We record the single counts of each SPD and the two-fold coincidences between SPD1 (SPD3) and SPD2 when $\Delta \lambda$ is changed. Moreover, for each measurement, we deduce true coincidence rate $C_T$ by subtracting coincidences measured from adjacent pulses from that of the same pulse. By measuring $C_T$ between SPD1 and SPD2, we check the probability of the $\left| \psi_1\right\rangle$ state; while by measuring $C_T$ between SPD2 and SPD3, we check the probability of the $\left| \psi_2\right\rangle$ state. The main plot and inset in Fig. 2(a) respectively show the counting rate of each SPD ($R_{s}$) and true coincidence rate $C_T$ versus $\Delta \lambda$. One sees the single counts of each SPD are about constant, and the data $C_T$ well fits the theoretical expectation. In particular, for the detuning of  $\Delta \lambda\approx10.75$ and $\Delta \lambda\approx15.2$\,nm, respectively, we have $\left| \psi_2\right\rangle$ and $\left| \psi_1\right\rangle$.

We then further check the contrast ratio of the switch by changing the pair production rate to demonstrate how well the SPFL can be used to route the signal and idler photons pairs to the desired spatial modes. We repeat the above measurements by changing the pump power for $\Delta \lambda=10.75$ and $\Delta \lambda=15.2$\,nm, respectively. The results show the contrast ratio does not vary with pump for both cases. As shown in Fig. 2(b) and 2(c), for $\Delta \lambda=10.75$, photon pairs in the same modes are negligible, the ratio between $C_T$ measured from the different and same spatial modes is about $100:1$; while for $\Delta \lambda=15.2$\,nm, photon pairs in different spatial modes are negligible, the ratio between $C_T$ measured from the same and different spatial modes is about $20:1$.  We think the smaller contrast ratio for $\Delta \lambda=15.2$\,nm is because the detuning is slightly deviated from the required ideal value.

In conclusion, we have presented a photon pair switch by exploiting the dispersion based phase modulation in SPFL.  When the photon pairs are used to generate heralded single photons~\cite{yang11}, the SPFL can function as a switch of single photons. Moreover, although a variety of fiber based nonlinear loops, inevitably embedded with the dispersive elements--standard single mode fibers, had been previously used to realize entangled photon pairs, the impact of the dispersive elements was ignored~\cite{Fiorentino02,Takesue04,Lim06,Chen07b,Chen08b,li09}. Therefore, our investigation is not only useful for quantum network, but also beneficial for designing all fiber sources of entangled photon pairs.

This work was supported in part by the NSF of China
(No. 11074186), the State Key
Development Program for Basic Research of China (No. 2010CB923101) and the Open Project of State Key Laboratory of Quantum Optics and Quantum Optics Devices.
$^{*}$ X. Li's e-mail address is xiaoyingli@tju.edu.cn.

\bibliographystyle{osajnl}


\end{document}